\documentclass[prl,amsmath,twocolumn,showpacs,superscriptaddress]{revtex4}
\usepackage{epsfig}

\newcommand{\noi}{\noindent}
\newcommand{\be}{\begin{equation}}
\newcommand{\ee}{\end{equation}}
\newcommand{\bdm}{\begin{displaymath}}
\newcommand{\edm}{\end{displaymath}}

\newcommand{\mbf}{\mathbf}

\begin{document}

\title{Generalized Elastic Model yields Fractional Langevin Equation}
                                                                                
\author{Alessandro Taloni}
\affiliation{School of Chemistry, Tel Aviv University, Tel Aviv 69978, Israel}
\author{Aleksei Chechkin}
\affiliation{School of Chemistry, Tel Aviv University, Tel Aviv 69978, Israel}
\affiliation{Akhiezer Institute for Theoretical Physics, NSC KIPT,
  Kharkov 61108, Ukraine}
\author{Joseph Klafter}
\affiliation{School of Chemistry, Tel Aviv University, Tel Aviv 69978, Israel}


\begin{abstract}
Starting from a generalized  elastic model which accounts
for the stochastic motion of several physical
systems such as membranes, (semi)flexible polymers and fluctuating
interfaces among others, we derive the 
fractional Langevin equation (FLE) for a probe particle in such 
systems, in the case of thermal initial conditions. We show that this
FLE is the only one   
fulfilling the fluctuation-dissipation (FD) 
relation within a new family of fractional Brownian motion (FBM)
equations. 
The FLE for the time-dependent fluctuations of the  
donor-acceptor distance in a protein, is shown to be recovered. When
the system starts from non-thermal conditions, 
the corresponding FLE, which does not fulfill FD relation, is derived.

\end{abstract}
\maketitle

%%%%%%%%%%%%%%%%%%%%%%%%%%%%%%%%%%%%%%%%%%%%%%%%%%%%%%%%%%%%%%%%%%%%%%%%%%%%%%%
%
%			I N T R O D U C T I O N
%
%%%%%%%%%%%%%%%%%%%%%%%%%%%%%%%%%%%%%%%%%%%%%%%%%%%%%%%%%%%%%%%%%%%%%%%%%%%%%%%
{\em Introduction.--} 
Continuum elastic models have been extensively used in statistical
mechanics to study the  dynamics of real physical
systems. Examples include: (semi)flexible
polymers ~\cite{Granek,semiflexible,Doi},
membranes ~\cite{membranes,membranes-FLE,Granek,Zilman}, growing
interfaces \cite{interfaces,Krug,Edwards,Gao,DeGennes}, fluctuating surfaces
\cite{surfaces} and diffusion-noise systems ~\cite{VKampen}. In
this paper we consider the following  Markovian equation for a
generalization of the accounted elastic models
(\emph{generalized elastic model}) 
\begin{equation}
\frac{\partial}{\partial t}\mathbf{h}\left(\vec{x},t\right)=\int d^dx'\Lambda\left(\vec{x}-\vec{x}'\right)\frac{\partial^z }{\partial\left|\vec{x}'\right|^z }\mathbf{h}(\vec{x}',t)+\boldsymbol\eta\left(\vec{x},t\right)
\label{Generalized_EW}
\end{equation}

\noi for the dynamics of the $D$-dimensional  stochastic process
$\mathbf{h}$ in the $d-$dimensional infinite space: the  internal $d$
coordinates  are represented by 
$\vec{x}$ and the Gaussian white noise satisfies the
fluctuation-dissipation (FD) relation
\be
\langle\eta_{j}\left(\vec{x},t\right)\eta_{k}\left(\vec{x}',t'\right) \rangle
=  2k_BT\Lambda\left(\vec{x}-\vec{x}'\right)\delta_{j\,k}\delta(t-t').
\label{FDT}
\ee
   
\noi with $j,k\in[1,D]$. The Riesz fractional operator  is
defined via its Fourier transform as ${\cal
  F}_{\vec{q}}\left\{\frac{\partial^z}{\partial\left|\vec{x}\right|^z}\right\}\equiv-\left|\vec{q}\right|^z$
($z>0$) ~\cite{Zazlawsky}
or, in terms of the Laplacian $\Delta$, as
$\frac{\partial^z}{\partial\left|\vec{x}\right|^z}:=-\left(-\Delta\right)^{z/2}$
~\cite{Samko}.
In the following  analysis we  study  two classes of
 hydrodynamic interactions: long ranged,
  $\Lambda\left(\vec{r}\right)\sim\left|\vec{r}\right|^{-\alpha}$ as $\vec{r}\to \infty$, with
 $0<\alpha<d$, and local, 
 $\Lambda\left(\vec{r}\right) = 
 \delta^d\left(\vec{r}\right)$. In the non-local case, if $\alpha=d$ we take
  $\Lambda\left(\vec{r}\right)\sim\frac{1}{a+\left|\vec{r}\right|^{d}}$
  where $a$ is a microscopic cutoff. 

\noi \emph{Long range hydrodynamic interactions.} The hydrodynamic
interactions are often represented by 
the equilibrium average of 
the Oseen tensor, which in an embedding $d_e$-dimensional space ($d_e\geq
3$) reads ~\cite{Doi}:
$\Lambda\left(\vec{r}\right)\sim\left|\vec{r}\right|^{2-d_e}$. Examples
are: (I)  \emph{fluid membranes} ~\cite{membranes,membranes-FLE,Granek,Zilman},
whose  height $h\left(\vec{x},t\right)$ of a point $\vec{x}$ on the
2-dimensional (planar) base surface is moving in time according to
(\ref{Generalized_EW}), with $\alpha=D+d-2=1$ and $z=4$ as derived from
the Helfrich bending free energy for small deformations ~\cite{Helfrich}. 
%for small deformations ($\vec{\nabla}h \ll 1$) ~\cite{Helfrich}.
(II) \emph{Semiflexible and flexible
  polymers' models}, where  $\mbf{h}$ represents the 3
spatial coordinates of a polymeric segment (\emph{bead}), while $x$ is
the strand's 1-dimensional internal coordinate (\emph{curvilinear
  abscissa}). For semiflexible filaments
~\cite{Granek,semiflexible} the
bending elastic energy associated with the chain's deformation 
implies $z=4$ ~\cite{Harris} and from the Oseen tensor formula we get 
$\alpha=D-2=1$; for the flexible polymers, often
referred to as \emph{Zimm model} ~\cite{Doi}, the free energy
contribution solely comes from the elastic term, i.e. $z=2$, and
$\alpha=1/2$ in $\Theta$ solvent.     

\noi \emph{Local hydrodynamic interactions.} Examples of the case
where hydrodynamic 
interactions are completely screened out are: (I) the \emph{Rouse model} 
equation  for polymer dynamics ~\cite{Doi}, once one sets $D=3$, $d=1$
and $z=2$. (II) \emph{Single file system}: recently 
it has been shown ~\cite{single_file} that the dynamics of a gas of
Brownian hard rods 
on a line  can be
mapped onto the harmonic chain problem ($z=2$), where $h(x,t)$ stands for the
position of the $x$-th particle on the 1-dimensional substrate at time
$t$.     (III) 
\emph{Fluctuating interfaces} ~\cite{interfaces,Krug}, where $h$ plays
the role of 
a scalar field (mostly the height of a rough surface in $d$ dimension) which is
subjected to a   
non-standard elastic force embodied by the fractional derivative of
order $z$. This is actually the generalization of the
Edwards-Wilkinson equation for the fluctuating profile of a granular
surface, for which $d=2$, $z=2$ ~\cite{Edwards}. In systems
such as crack propagation ~\cite{Gao} and contact line of a liquid
meniscus ~\cite{DeGennes} $d=1$ and the restoring forces are
characterized by $z=1$. If instead $h$ is 
meant to be a step, namely a line boundary at which   the surface
changes height by one or more atomic units, the value of $z$ in
eq.(\ref{Generalized_EW}) is 
found to be $z=2,3$ or $4$ ($d=1$) 
according to the character of the
atomic diffusion
%mechanism responsible for the fluctuations 
~\cite{surfaces}. (IV) \emph{Diffusion-noise equation}
~\cite{VKampen}: in this case $h$
represents the density field
on a $d$-dimensional surface  $\vec{x}$ and $z=2$. 
%(V) \emph{Signals}
%~\cite{signals}, for which eq.(\ref{Generalized_EW}) with $D=1$ and $d=1$ has
%been introduced to 
%account for the power-law spectra of fluctuations which scales with
%the frequency as $S(\omega)=1/\omega^z$.

The aim of this manuscript is to derive the
non-Markovian time-FLE for the field 
$h_j$ at a given position $\vec{x}$ (\emph{probe particle}), given that the
whole system's dynamics obeys the space fractional Markovian  equation
(\ref{Generalized_EW}). We interpret the FLE as a particular
case of a broader class of stochastic equations for FBM, i.e. the
generalized fractional Langevin equation.

%%%%%%%%%%%%%%%%%%%%%%%%%%%%%%%%%%%%%%%%%%%%%%%%%%%%%%%%%%%%%%%%%%%%%%%%%%%%%%%
%
%	      A U T O C O R R E L A T I O N  F U N C T I O N
%
%%%%%%%%%%%%%%%%%%%%%%%%%%%%%%%%%%%%%%%%%%%%%%%%%%%%%%%%%%%%%%%%%%%%%%%%%%%%%%%
\emph{Autocorrelation function.--}
We hereby calculate the
$h$-autocorrelation function $\langle
\delta h\left(\vec{x},t\right)\delta h\left(\vec{x},t'\right)
\rangle = \langle\left[
h_{j}\left(\vec{x},t\right)-h_{j}\left(\vec{x},0\right)\right]\left[h_{j}\left(\vec{x},t'\right)-h_{j}\left(\vec{x},0\right)
\right]\rangle$. Defining the 
Fourier transform in space and time as 
$h_j\left(\vec{q},\omega\right)=\int d^dx\, dt\,
e^{-i(\vec{q}\cdot\vec{x}-\omega t)}h_{j}\left(\vec{x},t\right)$,
the  solution of (\ref{Generalized_EW}) is 
\be
h_{j}\left(\vec{q},\omega\right)=\frac{\eta_{j}\left(\vec{q},\omega\right)}{-i\omega+\Lambda\left(\vec{q}\right)\left|\vec{q}\right|^{z}},
\label{sol_FF}
\ee

\noi where $\Lambda\left(\vec{q}\right)$ stands for  the $d$-dimensional
Fourier transform of 
the hydrodynamic interaction term, reading
$\Lambda\left(\vec{q}\right)=\frac{(4\pi)^{d/2}}{2^{\alpha}}\frac{\Gamma\left((d-\alpha)/2\right)}{\Gamma\left(\alpha/2\right)}\left|\vec{q}\right|^{\alpha-d}=A\left|\vec{q}\right|^{\alpha-d}$;
if  $\alpha=d$ in first
approximation we 
  can neglect the logarithmic corrections  in the
  hydrodynamic term's Fourier transform ~\cite{Granek}, i.e.
  $\Lambda\left(\vec{q}\right)\sim const$. In the local
hydrodynamic situation $\Lambda\left(\vec{q}\right)=1$, which
corresponds to put $A=1$
and $\alpha=d$ in the long-ranged hydrodynamic expression: this
substitution (which is not to be intended as a limit) allows us
to easily shift from power-law to local hydrodynamics throughout the
following analysis. Moreover, due to the isotropy of the problem under
study,  we
can drop the  
label $j$ in (\ref{sol_FF}).
% Since
%the Fourier transform of the noise 
%correlation function (\ref{FDT}) 
%gets the form  
%$\langle\eta\left(\vec{q},\omega\right)\eta\left(\vec{q}',\omega'\right) \rangle
%=2 k_BT
%(2\pi)^{d+1}\Lambda\left(\left|\vec{q}\right|\right)\delta\left(\omega+\omega'\right)\delta^d\left(\vec{q}+\vec{q}'\right)$, 
%after a bit of algebra we achieve that 
%\be
%\langle
%h\left(\vec{x},t\right)h\left(\vec{x},t'\right)
%\rangle  = k_BT
%\int_{-\infty}^{+\infty}\frac{d^dq}{(2\pi)^d}\frac{e^{-\Lambda\left(\vec{q}\right)\left|\vec{q}\right|^{z}\left|t-t'\right|}}{\left|\vec{q}\right|^{z}}.
%\label{EW_corr_func}
%\ee
%\noi The 
%explicit form of the
The $h$-autocorrelation function is readily obtained: the general
expression looks like 
\be
\langle
\delta h\left(\vec{x},t\right) \delta h\left(\vec{x},t'\right)
\rangle = 
K\left[t^{\beta}+t'^{\beta}-\left|t-t'\right|^{\beta}\right]
\label{EW_corr_func_general_form}
\ee

\noi where the anomlaous diffusion exponent $\beta$ and the diffusion
constant $K$ are defined as

\be
\left\{
\begin{array}{l}
\beta= \frac{z-d}{\alpha+z-d}\\ 
K  =  \frac{2k_BT\pi^{d/2}}{(2\pi)^d\Gamma\left(d/2\right)}\frac{A^{\beta}\Gamma\left(1-\beta\right)}{z-d}.
\label{EW_corr_func_z>d}
\end{array}
\right.
\ee

\noi Henceforth  we will concentrate on  the case $z>d$, which
does not need any 
regularization of the $\vec{q}$ integrals in the inverse Fourier
transform. The cases
$z<d$ and $z=d$ will be reported elsewhere. From
(\ref{EW_corr_func_general_form}) it is apparent that  the probe
particle placed at 
a given $\vec{x}$ performs 
a fractional Brownian motion (FBM) which is 
always subdiffusive, namely  $\langle\delta^2 h\left(\vec{x},t\right)\rangle=
2Kt^{\beta}$.

%%%%%%%%%%%%%%%%%%%%%%%%%%%%%%%%%%%%%%%%%%%%%%%%%%%%%%%%%%%%%%%%%%%%%%%%%%%%%%%
%
%	                               F L E
%
%%%%%%%%%%%%%%%%%%%%%%%%%%%%%%%%%%%%%%%%%%%%%%%%%%%%%%%%%%%%%%%%%%%%%%%%%%%%%%%
\begin{figure}
\begin{center}
\epsfig{figure=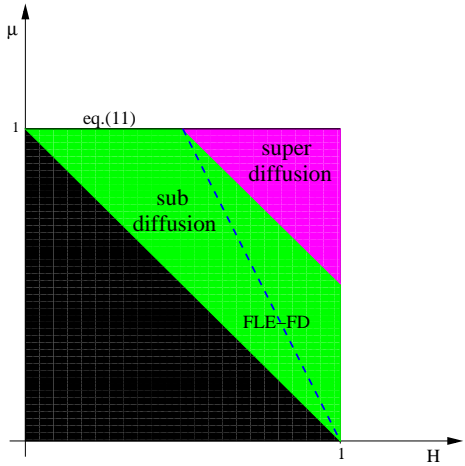, width=0.34\textwidth}
\end{center}
\caption{(Color online) Schematic picture of the GFLE ($\mu,H$) parameter
  space. 
%The upper black line 
%  corresponds to eq.(\ref{FBM}), while the dashed line  to
%  FLE fulfilling FD relation.
}
\label{fig1}
\end{figure}

\emph{Fractional Langevin Equation.--} We now proceed to derive the
fractional Langevin equation (FLE) for a probe particle placed at
position $\vec{x}$, which fulfills the FD relation. In
what follows we  show that this is the only fractional
stochastic equation physically relevant for the considered system
(\ref{Generalized_EW}).
We first include long-ranged hydrodynamic
interactions. 
We multiply both sides of the Fourier solution  
(\ref{sol_FF}) by $K^+(-i\omega)^{\beta}$, where
$K^+=\frac{k_BT}{K}\,\frac{1}{\Gamma\left(1+\beta\right)}.$
We then define the following function
\be
\Phi\left(\vec{x},\omega\right)=\int\frac{d^dq}{(2\pi)^d}\frac{e^{-i\vec{q}\cdot\vec{x}}\,(-i\omega)^{\beta}}{-i\omega+A\left|\vec{q}\right|^{z+\alpha-d}}.
\label{Phi}
\ee

\noi  Inverting the Fourier
transforms in space gives
\be
(-i\omega)^{\beta}K^+h\left(\vec{x},\omega\right)=\zeta\left(\vec{x},\omega\right)
\label{FLE_FT}
\ee

\noi where we have introduced the fractional Gaussian noise (fGN) 
$\zeta\left(\vec{x},\omega\right)=K^+\int d^dx'\,\eta\left(\vec{x}-\vec{x}',\omega\right)\Phi\left(\vec{x}',\omega\right)$. In
time domain the previous equation takes the final form of a FLE
~\cite{Lutz,single_file,Taloni,Kou,membranes-FLE}:
\be
K^+D_+^{\beta}h\left(\vec{x},t\right)=\zeta\left(\vec{x},t\right),
\label{FLE}
\ee

\noi where $D_+^{\beta}$ is the Caputo  derivative 
defined as  ~\cite{Samko,Podlubny}
\be
D_+^{\beta}f(t)=\frac{1}{\Gamma\left(1-\beta\right)}\int_{-\infty}^tdt'\frac{1}{\left(t-t'\right)^{\beta}}\frac{d}{dt'}f\left(t'\right),
\     \ 0<\beta<1.
\label{RL}
\ee

\noi The noise in (\ref{FLE}) can be shown to
satisfy the FD relation, 
\be
\langle \zeta\left(\vec{x},t\right)
\zeta\left(\vec{x},t'\right)\rangle=k_BT \frac{K^+}{\Gamma\left(1-\beta\right)\left|t-t'\right|^{\beta}}.
\label{FLE-FDT}
\ee

\noi In the absence of any hydrodynamic interaction the anomalous
diffusion exponent $\beta$ in
(\ref{FLE}) and (\ref{FLE-FDT}) is $\beta=1-d/z$ ~\cite{Krug}. Along
these lines, the FLE 
for the case 
 $d=1,\,z=2,\,\Lambda\left(x-x'\right)=\delta\left(x-x'\right)$ has
been obtained in ~\cite{single_file}: our derivation 
 can thus be viewed as a generalization of the technique used in such
 model. We emphasize that  the spatial
 correlations appearing in the model (\ref{Generalized_EW}) are
translated into  time correlations described by the
 fractional derivative (\ref{RL}), together with  the space-time
 correlations of the noise $\zeta\left(\vec{x},t\right)$. As a
 consequence, any $h$-correlation function can be
 calculated starting from (\ref{FLE}),
 e.g. $\langle\left[h\left(\vec{x},t\right)-h\left(\vec{x'},0\right)\right]^2\rangle$ 
 has been studied for fluctuating membranes ~\cite{Zilman}. We note
 that 
  the underlying assumption in the expression (\ref{FLE}) is that the
  system has reached the thermal equilibrium at $t=0$
  and
  the system's  
  configuration is drawn from the stationary Gibbsian probability distribution
  $\sim e^{-\frac{1}{2\,k_BT}\int{ d\vec{x}\left(\frac{\partial^{z/2}
    h\left(\vec{x},0\right)}{\partial\left|\vec{x}\right|
        ^{z/2}}\right)^2}}$ ~\cite{Santachiara}.

Now, it is possible to recast   the result (\ref{FLE-FDT}) in the
same fashion as the fGn  correlation function of a FBM
~\cite{Mandelbroot}: $\langle \zeta_H\left(t\right) 
\zeta_H\left(t'\right)\rangle\propto\left|t-t'\right|^{2H-2}$,
  with $H=1-\frac{\beta}{2}$. It stems from
  (\ref{EW_corr_func_general_form}) and (\ref{EW_corr_func_z>d}) 
that the $h$-autocorrelation function can be expressed as $\langle\delta h\left(\vec{x},t\right)\delta h\left(\vec{x},t'\right)
\rangle \propto t^{2-2H}+t'^{2-2H}-\left|t-t'\right|^{2-2H}$, which is
at odd  with the corresponding standard FBM quantity
~\cite{Mandelbroot}, whose the exponent 
is $2H$. However this is not surprising, since the two processes obey
two different stochastic fractional 
equations: \emph{i}) the FLE (\ref{FLE}) for  systems which fulfill
the FD relation, and \emph{ii}) the usual equation
\be
\frac{d B_H(t)}{dt}= \zeta_H\left(t\right)
\label{FBM}
\ee

\noi for FBM ~\cite{Mandelbroot}.
\begin{figure}
\begin{center}
\epsfig{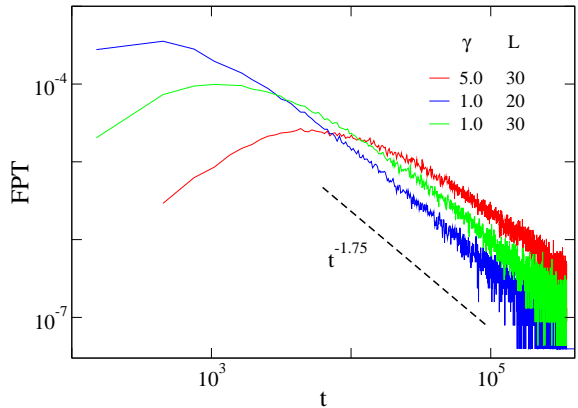}
\end{center}
\caption{(Color online) First passage time (FPT) distribution of the
  tagged single file particle initially placed at distance L from
  the adsorbing boundary. The system consists of $2\times 10^4$ Brownian
  particles subjected to hard-core interaction. Simulation' details
  can be found in ~\cite{single_file,Taloni}. The parameters
  are: the file's particle density $\rho=0.25$, the temperature
  $k_BT=1$ and the damping  $\gamma$. The theoretical
  prediction $\sim t^{-1.75}$ has been drawn for the reader
  convenience (dashed line). 
}
\label{fig2}
\end{figure}

%%%%%%%%%%%%%%%%%%%%%%%%%%%%%%%%%%%%%%%%%%%%%%%%%%%%%%%%%%%%%%%%%%%%%%%%%%%%%%%
%
%	                            G F B M
%
%%%%%%%%%%%%%%%%%%%%%%%%%%%%%%%%%%%%%%%%%%%%%%%%%%%%%%%%%%%%%%%%%%%%%%%%%%%%%%%
\emph{Generalized Fractional Langevin Equation.--} Let us now generalize eq.(\ref{FLE})
and (\ref{FBM}). Consider a stochastic process
$G(t)$ governed by the following dynamical equation
\be
D_+^{\mu}G(t)=\zeta_H\left(t\right)
\label{GFBM}
\ee

\noi where  $\zeta_H\left(t\right)$ is a fGn which satisfies $\langle
\zeta_H\left(t\right)\rangle=0$ and 
$\langle \zeta_H\left(t\right)
\zeta_H\left(t'\right)\rangle\simeq C\left|t-t'\right|^{2H-2}$ for
$\left|t-t'\right|\to\infty$ with $0<H<1$, $H\neq 1/2$;
$C<0$ if $0<H<1/2$, and $C>0$ if $1/2<H<1$. For $H=1/2$ ($C>0$)
$\zeta_H\left(t\right)$ is the white Gaussian noise ~\cite{Kobelev}. 
The fractional 
derivative has been defined in (\ref{RL}): it is 
immediate to recover  equation (\ref{FBM}) in the  limiting case
$\mu=1$, once that $D_+^{1}=  
d/dt$ and $C=2H(2H-1)$, and also the FLE (\ref{FLE}), setting
$K^+=1$, for systems satisfying the 
FD relation: $\mu=2-2H$ and
$C=k_BT/\Gamma(2H-1)$. The autocorrelation function can be calculated to yield
\begin{eqnarray}
&& \langle\delta
G\left(t\right)\delta G\left(t'\right)
\rangle=\frac{C\sin\left(\pi H\right)\Gamma\left(2H-1\right)}{\sin\left(\pi\left(H+\mu-1\right)\right)\Gamma\left(2H+2\mu-1\right)}\times\nonumber\\
&&\    \  \times \left[t^{2(H+\mu-1)}+t'^{2(H+\mu-1)}-\left|t-t'\right|^{2(H+\mu-1)}\right].
\label{GFBM_corr_func}
\end{eqnarray}

\noi Expression (\ref{GFBM_corr_func}) shows that $G(t)$ is a FBM with
Hurst exponent $H_{FBM}=H+\mu-1$ ($1<H+\mu<2$ ~\cite{footnote}). As a
consequence, 
systems for which $H+\mu<3/2$  exhibit  subdiffusive motion,
which instead is superdiffusive for  $H+\mu>3/2$. It is
interesting to note that the class of FBM for which the
FD relation holds can be \emph{only} subdiffusive.
These results are summarized in Fig.[\ref{fig1}].

\noi An important corollary of (\ref{GFBM_corr_func}) states that
any statistical property which is shown to be valid for a given pair
of values $\mu'$ and $H'$, is automatically valid for any other pair
($\mu'',\,H''$) which satisfies $\mu'+H'=\mu''+H''$. The
demonstration is 
straightforward: it is sufficient to note that, since $G(t)$ is
a Gaussian process, it is 
fully specified by the correlation function
(\ref{GFBM_corr_func}). As an example, take the first
passage time distribution (FPT) in a
semi-infinite domain for a FBM,  which 
decays asymptotically like $\sim t^{H_{FBM}-2}$
~\cite{Krug,Molchan}. We immediately get that the FPT distribution for a 
process which is solution of (\ref{GFBM}) is given by $\sim
t^{\mu+H-3}$, which in turn reads 
 $\sim t^{\frac{\beta}{2}-2}$ for systems obeying to
(\ref{Generalized_EW}). We numerically 
support this result as shown in Fig.[\ref{fig2}]. The FPT distribution
of a tagged particle in single file system, which has  been 
shown to be described by a FLE with $H=3/4$
~\cite{single_file,Taloni}, since $\mu=2-2H$,
attains the $\sim t^{-1.75}$ asymptotic behavior.

\noi On the other hand, given an FBM
process with $\langle\delta^2G(t)\rangle\sim t^{2H_{FBM}}$, there is no
chance to determine the correct pair $(\mu,H)$ among the
GFLEs (\ref{GFBM}) for which
$\mu+H-1=H_{FBM}$. Nevertheless, introducing 
an external potential does the trick: for instance, adding a
constant force $F$ on the 
right-hand side of (\ref{GFBM}) gives $\langle
G(t)\rangle=F\frac{t^{\mu}}{\Gamma(\mu+1)}$, while an harmonic force
$-\Omega^2 G(t)$ leads to the relaxation $\langle
G(t)\rangle/G(0)=\langle G(t)G(0)\rangle/\langle G^2(0)\rangle\to
E_{\mu,1}\left[-\left(t/t_0\right)^{\mu}\right]$, in case of deterministic
and thermal initial conditions respectively, where
$t_0=\Omega^{-2/\mu}$ and
$E_{\mu,\nu}$ stands for the Mittag-Leffler function
~\cite{Podlubny}. 
%For systems obeying to (\ref{Generalized_EW}) the
%$h$-autocorrelation function in presence of an harmonic potential is given
%by $\langle \delta h\left(\vec{x},t\right)\delta h\left(\vec{x},t'\right)\rangle/\langle \delta^2 h\left(\vec{x},0\right)\rangle\to
%E_{\frac{z-d}{z+\alpha-d},1}\left[\left(\right|t-t'\left|/t_0\right)^{\frac{z-d}{z+\alpha-d}}\right]$
%where $t_0=\left(\Omega^2/K^+\right)^{\frac{z+\alpha-d}{d-z}}$ ~\cite{membranes-FLE,Kou,single_file}.
%In general, however, given an anomalous diffusion Gaussian
%process, there is no chance to determine which GFBM (\ref{GFBM})
%(i.e. which couple of $\mu$ and $H$) is the more appropriate to
%reproduce its dynamics by merely looking at the mean square
%displacement. The only way to distinguish among the class of GFBM with
%constant $\mu+H-1$ is to add an external potential to the system and check for
%its moments. By instance, adding an harmonic force $-\Omega B_H(t)$ on
%the right-hand 
%side of (\ref{FBM}) gives the Mittag-Leffler decays $\langle
%B_H(t)\rangle/B_H(0)\equiv\langle 
%B_H(t)B_H(0)\rangle/\langle B_H^2(0)\rangle\to
%E_{\mu,1}\left[\left(t/\Omega^{\mu}\right)^{\mu}\right]$.  

\noi One
might question  the 
uniqueness of the FLE (\ref{FLE}) among the whole family of  GFLEs
for the probe particle $h\left(\vec{x},t\right)$. It is possible to show that
eq.(\ref{FLE}) is the unique GFLE  for a process
$h\left(\vec{x},t\right)$ whose dynamics is ruled  by
eq.(\ref{Generalized_EW}). The demonstration deals with the
introduction of a local constant force field
$F_j\delta\left(\vec{x}-\vec{x}^\star\right)\theta(t)$ on 
the right-hand side of eq.(\ref{Generalized_EW}). Since the system 
fulfills the FD relation (\ref{FDT}),  the connection
between the average drift of $h\left(\vec{x}^\star,t\right)$ and its
mean square displacement in the absence of force  is given by the
Einstein relation  
\be
\langle h\left(\vec{x}^\star,t\right)\rangle_F=F\frac{\langle \delta^2 h\left(\vec{x}^\star,t\right)\rangle}{2k_BT}.
\label{ER}
\ee 

\noi The only GFLE
which reproduces (\ref{ER}) is that which preserves the
FD relation, i.e. the FLE (\ref{FLE}). 

Let's now briefly discuss a practical example of the usefulness of
the framework developed here. In Refs~\cite{Kou,Min} Sunny Xie
and coworkers succeeded in modeling the motion of the donor-acceptor (D-A)
distance within a protein, as the coordinate of a fictitious particle
diffusing in an harmonic potential according to a FLE with
fractional derivative of order 1/2. In the spirit of
Refs~\cite{Tang,Debnath}, we consider an idealized Rouse chain 
as a model for the protein conformational dynamics. Therefore we take
$D=3,\, d=1,\, z=2$ and
$\Lambda\left(x-x'\right)=\delta\left(x-x'\right)$
in (\ref{Generalized_EW}). The D-A distance vector can be expressed as
$\mathbf{\Delta}_{D-A}(t)=\mathbf{h}\left(x_A,t\right)-\mathbf{h}\left(x_D,t\right)$,
and its correlation function by ~\cite{Debnath}
$\langle\mathbf{\Delta}_{D-A}(t)\cdot\mathbf{\Delta}_{D-A}(t')\rangle=3\langle\Delta_{D-A}(t)\Delta_{D-A}(t')\rangle$
due to the isotropy of the system. Hence, we can employ the result
of Ref.~\cite{single_file}, which shows that the generalized Langevin
equation for the single component $\Delta_{D-A}(t)$ is
$1/2\,D_+^{1/2}\Delta_{D-A}(t)=-\omega^2_0\left[\Delta_{D-A}(t)-\Delta_{D-A}(0)\right]+\zeta_{D-A}(t)$
in the long time limit , with  $\omega_0\propto(x_A-x_D)^{-1/2}$ and
$\zeta_{D-A}(t)$ satisfying the FD relation. 
However we point out here again that
$\langle\Delta_{D-A}(t)\Delta_{D-A}(t')\rangle$ can be evaluated
directly from eq.(\ref{FLE}).
%Our framework bridges in a straightforward way the Rouse
%model accounting for the dynamics of 
%the entire protein, to eq.(\ref{FLE}) for the 
%protein segment. 

%In Ref.~\cite{Tang} and
%~\cite{Debnath} the authors were able to reproduce the experimentally
%detected  D-A distance correlation function using a (respectively)
%discrete and continuous  microscopic Rouse model of the
%protein. This corresponds to take $d=1$, $z=2$ and $\Lambda\left(\vec{x}-\vec{x}'\right)=\delta^d\left(\vec{x}-\vec{x}'\right)$ in
%(\ref{Generalized_EW}) which, in turn, leads to the FLE of order 1/2
%in (\ref{FLE}). To complete the picture, in  Ref.~\cite{single_file}
%it has been shown 
%that the distance between two monomers (D-A) in the 1D Rouse model
%asymptotically obeys  the  1/2 FLE in the presence of a binding
%harmonic potential. We can thus assert that eq.(\ref{Generalized_EW})
%($z=2,\,d=1$) offers an extremely accurate dynamical representation of
%the entire protein, as well as eq.(\ref{FLE}) does for the single
%protein segment. However, the last is both theoretically and
%experimentally easier to treat.

%%%%%%%%%%%%%%%%%%%%%%%%%%%%%%%%%%%%%%%%%%%%%%%%%%%%%%%%%%%%%%%%%%%%%%%%%%%%%%%
%
%	          N O N - S T A T I O N A R Y  I N I T  C D T 
%
%%%%%%%%%%%%%%%%%%%%%%%%%%%%%%%%%%%%%%%%%%%%%%%%%%%%%%%%%%%%%%%%%%%%%%%%%%%%%%%
\emph{Non-thermal initial conditions.--} Let's now assume  that the initial
conditions for the system in (\ref{Generalized_EW}) are given by
\be
h\left(\vec{x},0\right)=0
\label{frozen_init}
\ee

\noi without
loss of generality. For systems such as
fluctuating interfaces ~\cite{interfaces,Krug,Edwards,surfaces} or membranes ~\cite{membranes,membranes-FLE,Granek,Zilman},
eq.(\ref{frozen_init}) assumes
the interface to  be flat at $t=0$. In the case of a
polymer, we can imagine eq.(\ref{frozen_init}) to be valid only
for the $j$-th 
component, achieving an initial configuration which is randomly
arranged within the plane $j=0$. For single file systems eq.(\ref{frozen_init})
consists of taking  particles equally spaced at $t=0$.  The
$h$-autocorrelation 
function can be obtained in the same fashion as in the case of thermal
initial conditions by using Laplace transform in time instead of the
Fourier transform.  A straightforward calculation
yields

\be
\langle
\delta h\left(\vec{x},t\right) \delta h\left(\vec{x},t'\right)
\rangle = 
K\left[\left(t+t'\right)^{\beta}-\left|t-t'\right|^{\beta}\right]
\label{EW_corr_func_frozen_general_form}
\ee

\noi where the value of $K$ and $\beta$ get the same expressions as in
%(\ref{EW_corr_func_z<d}), (\ref{EW_corr_func_z=d}),
(\ref{EW_corr_func_z>d}). For local hydrodynamics
eq.(\ref{EW_corr_func_frozen_general_form}) 
matches the result 
previously obtained  by Krug \emph{et al.} for
fluctuating interfaces ~\cite{Krug}.

\noi It is easy to show that the FLE expression (\ref{FLE}) is
still valid, with the Caputo derivative having its lower terminal at
$t=0$ ~\cite{Caputo,Podlubny}. When
attempting to 
recover the FD 
relation, however, one gets the following form of the noise
correlation function
\be
\langle \zeta\left(\vec{x},t\right)
\zeta\left(\vec{x},t'\right)\rangle=\frac{k_BT\,K^+}{\Gamma\left(1-\beta
  \right)}\left[\left|t-t'\right|^{-\beta}-\left(t+t'\right)^{-\beta}\right].
\label{FLE-FDT_frozen}
\ee

\noi Expression (\ref{FLE-FDT_frozen}) clearly shows that the
noise $\zeta$ attains the stationarity (\ref{FLE-FDT})
in the limit ($t,t'\to\infty$).

%%%%%%%%%%%%%%%%%%%%%%%%%%%%%%%%%%%%%%%%%%%%%%%%%%%%%%%%%%%%%%%%%%%%%%%%%%%%%%%
%
%	              D I S C U S S I O N 
%
%%%%%%%%%%%%%%%%%%%%%%%%%%%%%%%%%%%%%%%%%%%%%%%%%%%%%%%%%%%%%%%%%%%%%%%%%%%%%%%
\emph{Discussion.--} In this Letter we presented the
derivation of the FLE for a wide class of phenomena, whose 
stochastic dynamics is ruled by the generalized elastic model
(\ref{Generalized_EW}).  The introduced
framework offers theoretical and practical advantages. On one hand, different
physical systems can be defined on the basis of a unique index: the
fractional derivative order (\emph{universality class}). On the other
hand, the FLE allows to achieve 
the relevant statistical observable by simply solve/simulate a non-Markovian
linear equation for the probe particle. Finally, from an
experimental perspective, the FLE description  allows the straightforward
detection of the microscopical parameters characterizing the system
(\ref{Generalized_EW}). 
%e.g. the protein's length,  degree of stiffness, and
%the D-A distance along the chain, in the aforementioned protein
%experiment ~\cite{Kou,Min}.

A.C. and J.K. acknowledge the support of Marie Curie IIF programme,
grant ``LeFrac''. A.T. thanks A. Rosso, M. Lomholt, L. Lizana,
T. Ambj\"ornsson  and R. Granek for valuable comments.


\begin{thebibliography}{99}
\bibitem{footnote} The case $1<\mu<2$ will be studied in a following
  publication.
\end{thebibliography}
\end{document}